# Pressure-induced Superconductivity in the Iron-based Ladder Material BaFe$_2$S$_3$

Hiroki Takahashi[1], Akira Sugimoto[1], Yusuke Nambu[2,3], Touru Yamauchi[4], Yasuyuki Hirata[4], Takateru Kawakami[1], Maxim Avdeev[5,6], Kazuyuki Matsubayashi[4], Fei Du[4,7], Chizuru Kawashima[1], Hideto Soeda[1], Satoshi Nakano[8], Yoshiya Uwatoko[4], Yutaka Ueda[4,9], Taku J. Sato[2], Kenya Ohgushi[4,10]

1. College of Humanities & Sciences, Nihon University, Sakurajosui, Setagaya-ku, Tokyo 156-8550, Japan
2. Institute of Multidisciplinary Research for Advanced Materials, Tohoku University, Sendai, Miyagi 980-8577, Japan
3. Institute for Materials Research, Tohoku University, Sendai, Miyagi 980-8577, Japan
4. Institute for Solid State Physics (ISSP), University of Tokyo, Kashiwa, Chiba 277-8581, Japan
5. Bragg Institute, Australian Nuclear Science and Technology Organisation, Locked Bag 2001, Kirrawee DC NSW 2232, Australia
6. School of Chemistry, The University of Sydney, Sydney, NSW 2006, Australia
7. Key laboratory of Physics and Technology for Advanced Batteries (Ministry of Education), College of Physics, Jilin University, Changchun 130012, People's Republic of China
8. Ultra-High Pressure Processes Group, National Institute for Materials Science, Tsukuba, Ibaraki 305-0044, Japan
9. Toyota Physical and Chemical Research Institute, Nagakute, Aichi 480-1192, Japan
10. Department of Physics, Tohoku University, Sendai, Miyagi 980-8578, Japan

*hiroki@chs.nihon-u.ac.jp, *ohgushi@m.tohoku.ac.jp

**All the iron-based superconductors identified to date share a square lattice composed of Fe atoms as a common feature, despite having different crystal structures. In copper-based materials, the superconducting phase emerges not only in square lattice structures but also in ladder structures. Yet iron-based superconductors without a square lattice motif have not been found despite being actively sought out. Here, we report the discovery of pressure-induced superconductivity in the iron-based spin-ladder material BaFe$_2$S$_3$, a Mott insulator with striped-type magnetic ordering below ~120 K. On the application of pressure this compound exhibits a metal-insulator transition at about 11 GPa, followed by the appearance of superconductivity below $T_c$ = 14 K, right after the onset of the metallic phase. Our findings indicate that iron-based ladder compounds represent promising material platforms, in particular for studying the fundamentals of iron-based superconductivity.**

The discovery of iron-based superconductors had a significant impact on condensed matter physics and led to extensive study of the interplay between crystal structure, magnetism and superconductivity[1]. All the iron-based superconducting materials discovered to date share the same structural motif: a two-dimensional square lattice formed by edge-shared Fe$X_4$ tetrahedra ($X$ = Se, P and As). The Fe atoms are nominally divalent in most of the parent materials. These parent compounds undergo a magnetic transition at low temperatures, typically exhibiting striped-type ordering. Superconductivity appears when the magnetic order is fully suppressed by the application of pressure or by the addition of doping carriers through chemical substitution. Based on an itinerant model, the striped-type magnetic ordering is stabilized by Fermi surface nesting and the associated antiferromagnetic fluctuations induce a superconducting state with so-called $s^{\pm}$ wave symmetry[2-4]. However, there is still significant disagreement concerning the degree of electron correlation in iron-based superconductors[5], sparked by the discovery of the 245 system, $A_2$Fe$_4$Se$_5$ ($A$ = K, Rb and Cs)[6-7], in which superconductivity emerges in the vicinity of the Mott insulating state with block-type antiferromagnetic ordering.

**Iron-based ladder materials**
$A$Fe$_2$Se$_3$ ($A$= Ba, K and Cs) crystallizes in a quasi-one-dimensional structure consisting of ladders formed by edge-shared FeSe$_4$ tetrahedra with channels occupied by $A$ atoms, as shown in Fig. 1a[8-16]. These compounds all exhibit unique magnetic ordering. The magnetic structure of $BaFe_2Se_3$ is block-type, in which the magnetic moments form $Fe_4$ ferromagnetic units which stack antiferromagnetically along the leg direction of the ladder[12-14] in a one-dimensional analogue of the block magnetism observed in the 245 system. In contrast, the magnetic structures of the $AFe_2Se_3$ ($A$ = K and Cs) are of the stripe-type, in which the magnetic moments couple ferromagnetically along the rung and antiferromagnetically along the leg direction[13-16]. This magnetic structure represents a one-dimensional analogue of the stripe magnetism observed in 1111 (e.g. LaFeAsO) and 122 (e.g. $BaFe_2As_2$) systems. Since these spin-ladder materials exhibit various magnetic structures similar to those of the iron-based superconductors, they can themselves be considered as potential superconductors. To date, however, they have shown only insulating characteristics with regard to their electrical resistivity, and have not yet been demonstrated to exhibit a metallic state.

**Purpose of this study**

The application of pressure is often a useful means of changing the electronic structure of a compound so as to induce a metallic state without simultaneously introducing any degree of disorder[17]. In this study, we investigated in detail the magnetic properties of a sulphur-analogue of the Fe-based ladder materials, $BaFe_2S_3$ (space group: orthorhombic, *Cmcm*) [18,19], and undertook experimental trials in which this compound was subjected to high pressures to obtain the metallic state. The electronic properties of this material depend on the manner in which the samples are synthesized, and thus we present data for sample 1 describing magnetic properties, and data for a range of samples 1 to 6 describing high-pressure effects. The details of the sample preparation process are given in the Method section.

**Electronic properties under ambient pressure**

Figure 2a displays the temperature dependence of the electrical resistivity ($\rho$) of $BaFe_2S_3$ along the leg direction under ambient pressure. The observed insulating behaviour, which occurs despite the expected metallic behaviour in an unfilled 3*d* manifold, is caused by the Coulomb repulsion between Fe 3*d* electrons, which becomes prominent in a quasi-one-dimensional ladder structure. Figure 2b shows the magnetic susceptibility ($\chi$) at 5 T along the three orthorhombic principal axes. These $\chi$ curves show a decrease in magnitude on cooling, which markedly contrasts to the Curie-Weiss behaviour typically observed in localized spin systems. This unique feature is reminiscent of the $\chi$ curves of the 1111 and 122 systems of iron-based superconductors[1,20], and is possibly caused by a gradual evolution of antiferromagnetic correlations. On further cooling, anisotropy in $\chi$ emerges below $T_N$ = 119 K, indicating long-range antiferromagnetic ordering. The smaller $\chi$ values found along the *a*-axis compared with $\chi$ values along the other two axes is persuasive evidence that the antiferromagnetic magnetic moments are aligned along the rung direction.

**Magnetic structure**

To examine the magnetic structure below $T_N$, high-resolution powder neutron diffraction measurements were carried out. The details of the determination of magnetic structure are provided in the Supplementary Information (Supplementary Fig. 1, Tables 1 and 2). The optimized refinement of the magnetic structure indicates a stripe-type arrangement, with the magnetic structure depicted in Figs. 1b and 1c. The estimated moment at 4 K is 1.20(6) $\mu_B$ per Fe site, which is smaller than that of the fully localized case, 4 $\mu_B$; such feature is also observed in iron-based superconductors. These magnetic moments are arranged to form ferromagnetic units along the rung direction, stacking antiferromagnetically along the ladder direction. For inter-ladder coupling on the *ab*-plane, a ferromagnetic spin unit correlates antiferromagnetically with one neighbouring ladder and ferromagnetically in the other direction. The single-stripe magnetic structure is reminiscent of the 1111 and 122 systems found in the iron-based superconductors[1,20]. The ladder-type $A$Fe$_2$Se$_3$ ($A$ = K and Cs) also possess stripe-type structures, although in the case of these materials the moment direction is instead parallel to the leg[13,16]. On warming, the magnetic moment disappears at 119 K as shown in Fig. 2c, which corresponds to $T_N$ determined by the magnetic susceptibility measurements shown in Fig. 2b. The nature of the magnetic phase transition is discussed on the basis of the specific heat and the Mössbauer spectra in the Supplementary Information (Supplementary Figs 2 and 3).

**Pressure-induced superconductivity**

Intrigued by the fairly small electrical resistivity shown by BaFe$_2$S$_3$ even in the Mott insulating state, together with the resemblance of its magnetic structure to that of iron-based superconductors, we assessed its superconductivity under high pressure generated in a diamond anvil cell (DAC) and a cubic anvil press. Figures 3a and 3b show the resistance (*R*) values obtained under pressures up to 13 GPa for sample 1 using DAC. On the application of pressure, the insulating characteristics are gradually suppressed and the *R* curve exhibits a metal-insulator transition between 10 and 11 GPa. The *R* plot at 11 GPa clearly shows a sudden decrease in resistance around 13 K and this feature is attributed to a superconducting transition, based on two observations. Firstly, as shown in Fig. 3c, the $R$ curve at 11.5 GPa shifts to the low-temperature side on the application of a magnetic field up to 7 T: the $T_c$ values as a function of the magnetic field ($H$) is shown in the inset of Fig. 3c, where $dT_c/dH \sim -3.8$ K/T. Secondly, the $R$ curve also shifts to the low-temperature side with increases in the electric current, as seen in Fig. 3d. The lack of a measurement of zero resistance could be due to technical limitations inherent in the experimental apparatus, since fully symmetric hydrostatic compressive stress could not be applied inside the DAC when using a solid pressure-transmitting medium. The value of $T_c$ was determined as illustrated in Fig. 3b. High-pressure experimental trials using DAC were performed once for samples 2, 3 and 5 and twice for samples 1 and 4, and these confirmed that the onset of superconductivity is reproducible, with the exception of one trial for sample 2(Supplementary Figs 4 and 5).In addition, Fig. 3e shows the $R$ plot for sample 6 at 13.5 GPa measured using cubic anvil press and glycerine as the liquid pressure-transmitting medium. Zero resistance is clearly observed below about 15 K. Fig. 3f shows ac susceptibility for sample 5 at 14.5 GPa. The diamagnetic signal is observed below about 13 K. Based on the diamagnetic signal from Pb which is observed below 5 K, the shielding volume fraction of $BaFe_2S_3$ is about 64% at 5 K. These data from the measurements in the cubic anvil press indicate that this superconductivity has a bulk origin. The pressure dependencies of $T_c$ for all the samples are summarized in Fig. 4b, in which $T_c$ is seen to describe a dome shape with a maximum $T_c$ of 17 K at 13.5 GPa.

**Crystal structure under high pressure**
We subsequently investigated the crystal structures under high pressure to acquire additional information concerning the microscopic mechanisms of the metal-insulator transition and the appearance of superconductivity. The X-ray powder diffraction patterns acquired under pressure did not show definite signs of structural transition (Supplementary Fig. 6), indicating that the superconductivity appears in the ladder structure. Tiny impurity peaks designated by * symbols are seen in the powder X-ray diffraction patterns (Supplementary Fig. 6), which may be attributed to unknown inclusions inside the single crystal. The presence of superconducting FeSe can be ruled out, because the impurity peaks do not match those of FeSe. Also, the large volume fraction of the superconducting phase from the ac susceptibility measurements demonstrates that the superconductivity of $BaFe_2S_3$ is of bulk origin. Figure 4a shows the pressure dependencies of lattice constants at 300 K, normalized to the ambient pressure values. Here the $b$ axis perpendicular to the ladder layer is seen to be the most compressible and it is likely that anisotropic compression makes the system more metallic by increasing transfer of the Fe $3d$ electrons across the inter-ladder bonds. In this sense, the observed metal-insulator transition is categorized as a bandwidth-control-type Mott transition. We could not acquire information concerning the magnetic ordering under high pressures although, considering the general trend in the iron-based and copper-based superconductors in which magnetic ordering is suppressed with the application of pressure, it is reasonable to assume that the increased conductivity disturbs the magnetic ordering in $BaFe_2S_3$. The superconductivity likely emerges in the vicinity of the quantum critical point associated with both charge and spin fluctuations.

**Comparison with copper-based ladder superconductors**
It is clear that $BaFe_2S_3$ represents a new series of iron-based superconductors, even though the anionic element, S, differs from the elements Se, P and As found in the known iron-based superconductors. Our finding of superconductivity on the verge of the Mott transition is additional evidence of the important role that the electron correlation effect plays important roles in iron-based superconductors and should stimulate further theoretical analysis[21-25]. The $d$ wave symmetry of the superconductivity is theoretically predicted in the ladder compounds, and is distinct from the common $s^{\pm}$ wave symmetry of the two-dimensional square lattice structures[22]. From the structural point of view, $BaFe_2S_3$ is quite similar to the copper-based ladder material $Sr_{14-x}Ca_xCu_{24}O_{41}$[26-29]. In this compound, the isovalent substitution of Ca at Sr sites in conjunction with the application of pressure transfers the charge from the CuO chain to the $Cu_2O_3$ ladder, resulting in suppression of charge ordering and the appearance of superconductivity . In this sense, the metal-insulator transition in $Sr_{14-x}Ca_xCu_{24}O_{41}$ is categorized as a filling-control-type Mott transition, contrasting to the bandwidth-control-type Mott transition in $BaFe_2S_3$. Another difference is that even though no definite magnetic ordering has been observed adjacent to the superconducting phase in $Sr_{14-x}Ca_xCu_{24}O_{41}$, the superconductivity emerges adjacent to the magnetic phase in $BaFe_2S_3$. We conclude that the superconductivity in $BaFe_2S_3$ bridges the behaviour of various unconventional superconductors, offering an ideal platform for the study of the interplay between magnetism and superconductivity. Further chemical modification, e.g. applying "chemical pressure" may be expected to shift the transition pressure to much more accessible range, possibly to ambient pressure.

Acknowledgements

This work was partly supported by JSPS Grants-in-Aid for Scientific Research (A) (23244068), for Scientific Research (B) (23340097, 24340088 and 26287073) and for Young Scientists (B) (26800175), the Strategic Research Base Development Program for Private Universities (2009, S0901022) of MEXT and the Grant Program of the Yamada Science Foundation.

**Author Contributions**

H.T., Y. N. and K. O. designed the research, Y. H., F. D., Y. U. and K. O. synthesized the samples and performed the resistivity, magnetic susceptibility and specific heat measurements, M. A., Y. N. and T. S. performed the neutron diffraction measurements and analysed the data, T. K. performed the Mössbauer measurements, A. S., C. K., H. S. and H. T. performed the resistivity measurements under high pressure at zero magnetic field, and A. S., C. K, H. T., K. M. and Y. U. performed the resistivity measurements under high pressure at finite magnetic fields. T.Y. performed the resistivity and magnetic susceptibility measurements under high pressure using the cubic anvil press, and C. K., S. N. and H. T. performed the X-ray diffraction measurements under high pressure. All the authors discussed the results, H. T., Y. N., T. K. and K. O. wrote the paper and all the authors read and commented on the manuscript.

**Author Information**

**Correspondence and requests for materials should be addressed to H.T. (hiroki@chs.nihon-u.ac.jp) or K.O. (ohgushi@m.tohoku.ac.jp).**

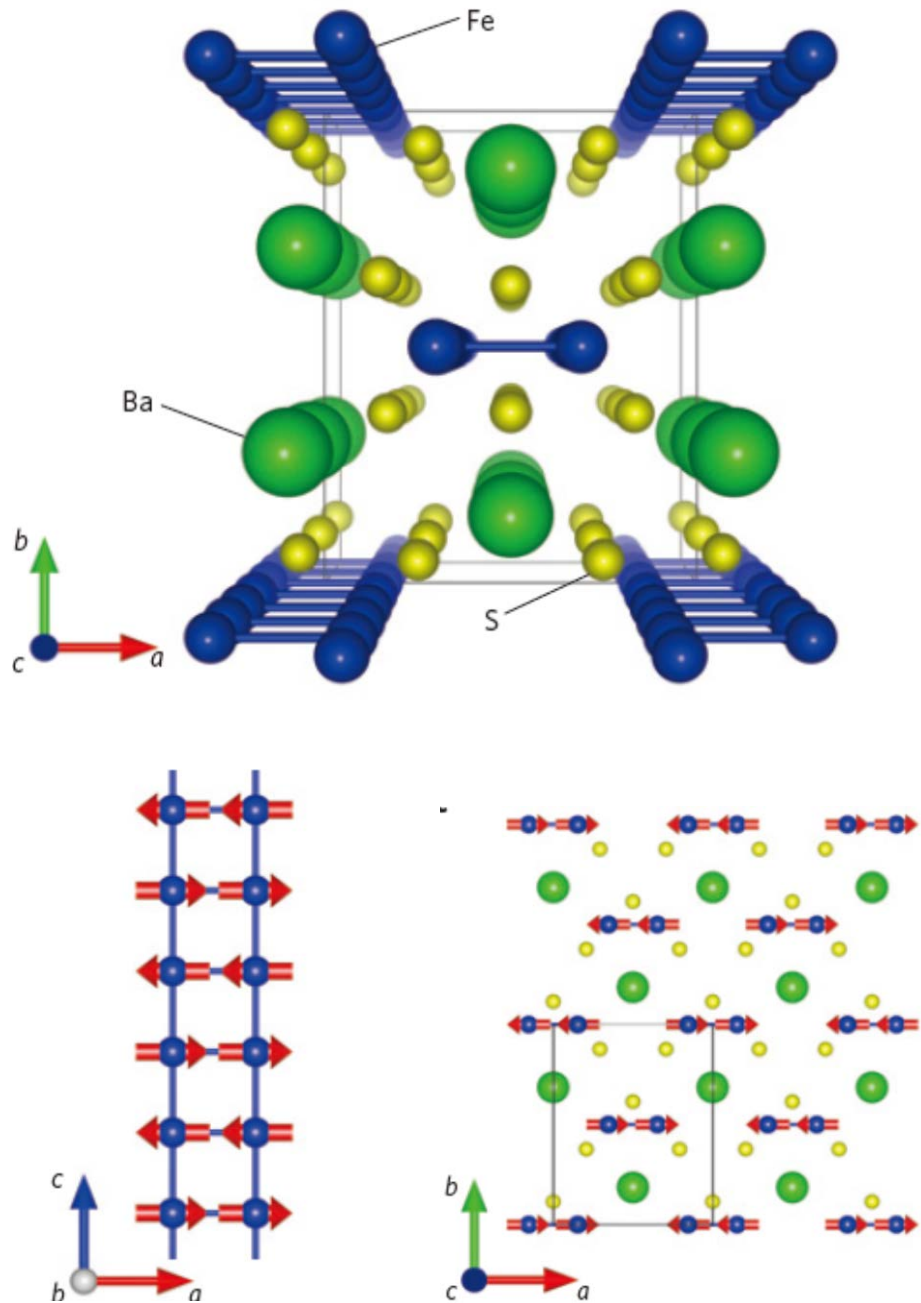

**Figure 1. Schematic view of the crystallographic and magnetic structures of BaFe$_2$S$_3$. a.** The crystal structure of BaFe$_2$S$_3$, consisting of edge-shared FeS$_4$ tetrahedra extending along the *c*-axis and channels occupied by Ba atoms, in which Fe atoms form a ladder structure. **b.** The magnetic structure in the ladder. Magnetic moments pointing to the *a*-axis form ferromagnetic units along the rung direction and stack antiferromagnetically along the leg direction. **c.** Magnetic structure viewed from the *c*-axis. The inter-ladder coupling of Fe spins is ferromagnetic along the 1 1 0 direction and antiferromagnetic along the 1 -1 0 direction.

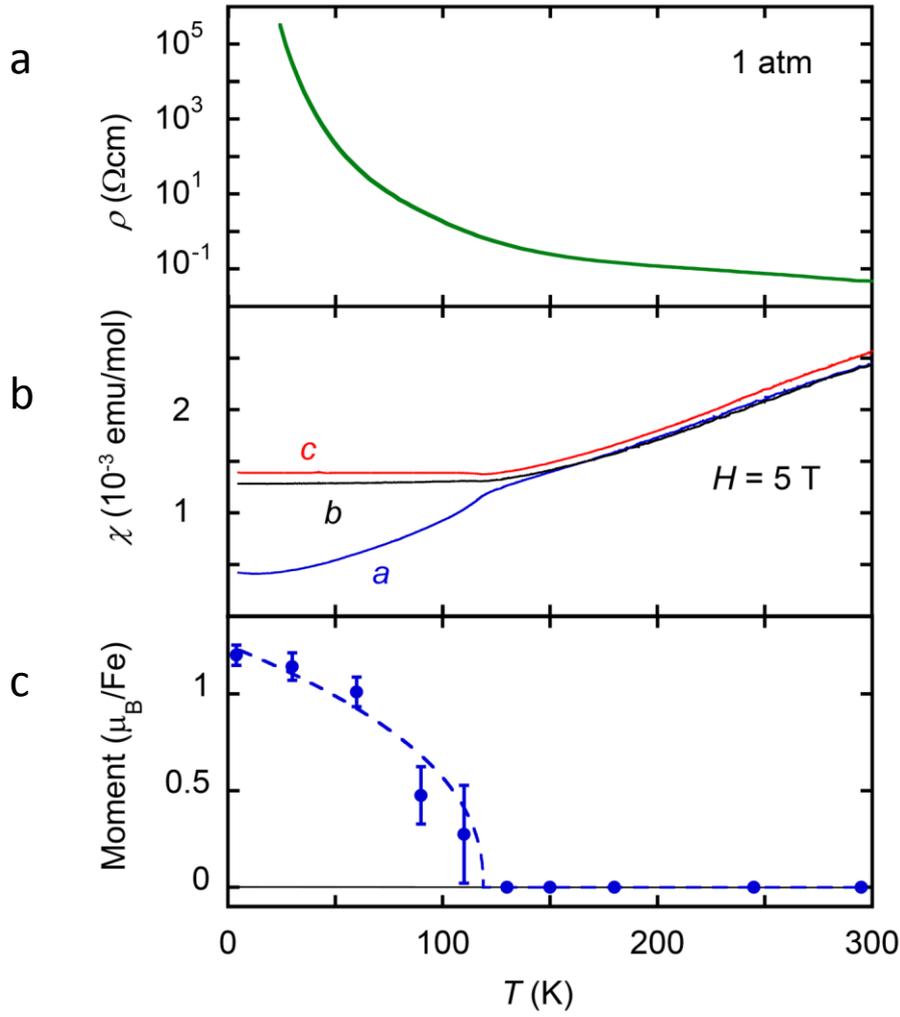

**Figure 2. Electronic properties of BaFe$_2$S$_3$ at ambient pressure. a.** Temperature ($T$) dependence of electrical resistivity ($\rho$), demonstrating insulating behaviour. **b.** Temperature dependence of magnetic susceptibility ($\chi$) along the three principle axes under a magnetic field of 5 T, demonstrating antiferromagnetic transition at $T_\text{N}$ = 119 K. **c.** Temperature dependence of the magnetic moments obtained from neutron diffraction measurements. Error bars are estimated from the standard deviation of moment in Rietveld refinement.

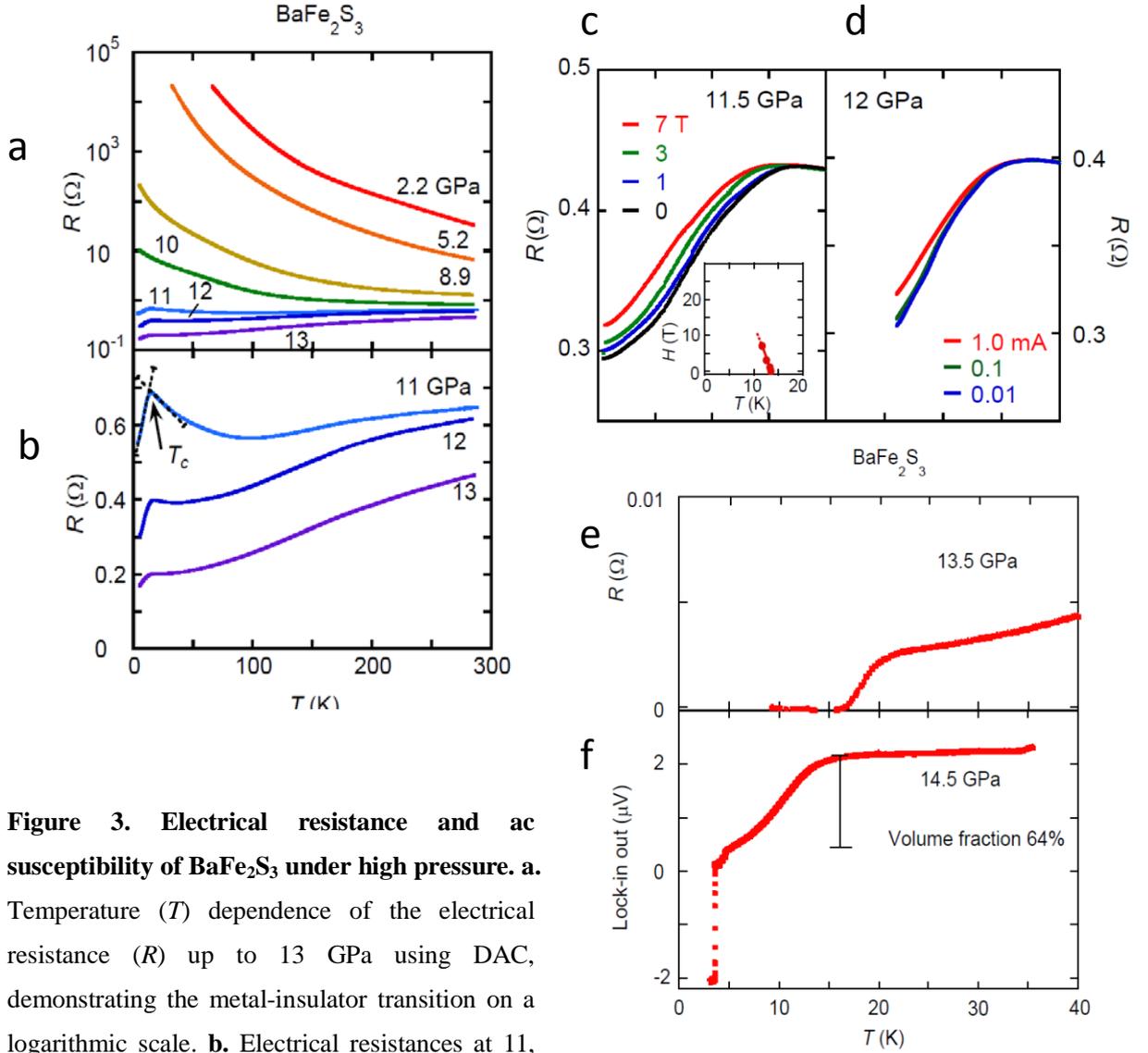

**Figure 3. Electrical resistance and ac susceptibility of BaFe$_2$S$_3$ under high pressure. a.** Temperature ($T$) dependence of the electrical resistance ($R$) up to 13 GPa using DAC, demonstrating the metal-insulator transition on a logarithmic scale. **b.** Electrical resistances at 11, 12 and 13 GPa on a linear scale, showing a superconducting transition. **c.** Electrical resistances at 11.5 GPa under various magnetic fields. The inset shows the magnetic field ($H$) dependence of $T_c$. **d.** Electrical resistance at 12 GPa as a function of the current. **e.** $T$ dependence of $R$ at 13.5 GPa measured using cubic anvil press, exhibiting a zero resistance below about 15 K caused by superconducting transition. **f.** $T$ dependence of the ac susceptibility at 14.5 GPa, showing diamagnetic signal below about 13 K. The diamagnetic signal from Pb is also observed below 5 K. Based on the diamagnetic signal from Pb, the shielding volume fraction of BaFe$_2$S$_3$ is about 64 % at 5 K, which clearly indicates that this superconductivity has a bulk origin.

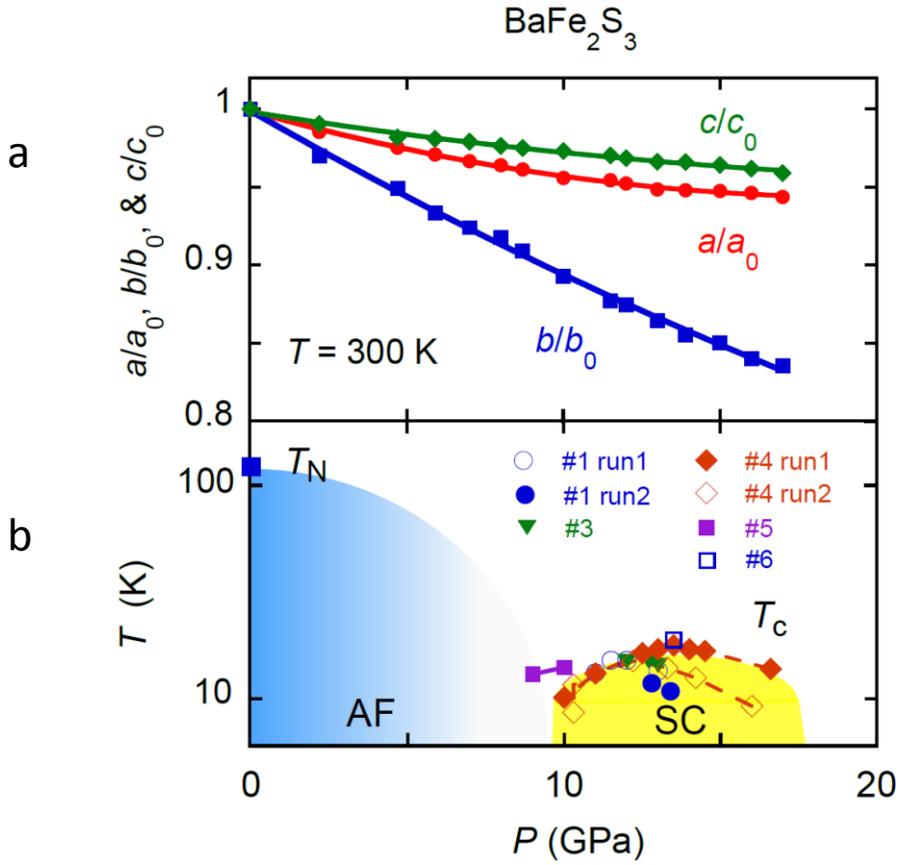

**Figure 4. Pressure dependence of the lattice constants and the superconducting transition temperature. a.** Pressure ($P$) dependence of lattice constants normalized to the ambient pressure values, obtained by high-pressure X-ray diffraction. The layer direction ($b$-axis) is seen to be the most compressible. **b.** The electronic phase diagram in the pressure-temperature ($P$-$T$) plane, including the superconducting (SC) transition temperatures ($T_c$) of all samples. The $T_c$ of "#6" was obtained by the resistivity measured using cubic anvil press. $T_N$ indicates the antiferromagnetic (AF) transition temperature, expected to be suppressed by the application of pressure.

**Methods**

**Sample preparation.** Single crystals of $BaFe_2S_3$ were synthesized by the solid state reaction method. The starting materials and their ratios differed between sample batches as follows: batches 1 and 3: Ba:Fe:S = 1:2:3; batch 2: BaS:Fe:S = 1:2:2; and batches 4, 5 and 6: BaS:Fe:S = 1:2.1:2. Commercially available BaS was used in the synthesis of batches 2 and 4, while BaS made from Ba and S at 700 °C within a carbon crucible inside a quartz tube was used in the synthesis of batch 5 and 6. In each case, the starting materials, at a total combined mass of approximately 1.5 g, were placed into the carbon crucible, following which the crucible was sealed inside the quartz tube under Ar at 0.3 atm. The tube was subsequently heated to 1100 °C over a period ranging from 8 to 40 h, maintained at that temperature for 24 h, and then slowly cooled to 750 °C over 24 h. The resulting crystals exhibited needle-like morphologies with the needle axis along the leg direction. Crystals were crushed prior to characterization by powder X-ray diffraction with Cu $K\alpha$ radiation.

**Characterization.** The resistivity of each sample along the leg direction was measured using the standard four-probe method, applying gold paste as an electrode. Magnetic susceptibility data were collected employing a superconducting quantum interference device (SQUID) magnetometer. Heat capacity measurements were performed using a commercial instrument and applying the relaxation method.

**Mössbauer spectroscopy.** $^{57}$Fe Mössbauer measurements at temperatures as low as 9 K were acquired for crushed crystals in Rh matrixes, using a 925 MBq $^{57}$Co γ-ray source with an active area 5 mm in diameter. The velocity scale of the spectra was relative to that of $\alpha$-Fe at room temperature.

**Neutron diffraction.** Neutron powder diffraction data were collected on the high-resolution ECHIDNA diffractometer at the Australian Nuclear Science and Technology Organisation (ANSTO) using an incident neutron wavelength of 2.4395(2) Å. Diffraction patterns were obtained between 4 and 295 K in a closed-cycle refrigerator.

**High-pressure experiments.** High-pressure resistivity measurements were carried out by a standard dc four-probe method using a diamond anvil cell (DAC) and a cubic anvil press. A DAC made of CuBe alloy was used for electrical resistance measurements at pressures up to 30 GPa. The sample chamber was filled with powdered NaCl as the pressure-transmitting medium, using a rhenium gasket, and thin (10 μm thick) platinum ribbons were used as leads for the standard dc four-probe analysis. Each rectangular sample was 0.1 mm × 0.1 mm and 0.03 mm thick. A thin BN layer acted as electric insulation between the leads and the rhenium gasket and finely ground ruby powder scattered in the sample chamber was used to determine the pressure by the standard ruby fluorescence method. A cubic anvil press equipped with sintered diamond anvils was used up to 14.5 GPa. Pressure was retained nearly hydrostatic in a Teflon capsule with a liquid transmitting medium, glycerine. The Teflon cell was placed in the cubic type gasket (4 mm × 4 mm × 4mm) made of MgO. Gold wires was used as electrical leads with gold paint contact on the surface of sample and connected to the anvil surface through thin gold ribbon. A cubic anvil press was also used for ac susceptibility measurements. Primary and secondary coils were wounded around the sample and Pb in the Teflon capsule. The high-pressure X-ray diffraction measurements were performed using synchrotron radiation at PF-BL18C at the High Energy Accelerator Research Organization (KEK), applying a wavelength of 0.621 Å to a sample situated in a DAC with a liquid pressure-transmitting medium (Daphene 7474). To check the bulk superconductivity, electrical resistance and magnetic susceptibility measurements were performed using the cubic anvil press, in which the sample chamber was filled with glycerine as the pressure-transmitting medium.

**Supplementary Information for "Pressure-induced Superconductivity in the Iron-based Ladder Material BaFe$_2$S$_3$"**

Hiroki Takahashi[1], Akira Sugimoto[1], Yusuke Nambu[2,3], Touru Yamauchi[4], Yasuyuki Hirata[4], Takateru Kawakami[1], Maxim Avdeev[5,6], Kazuyuki Matsubayashi[4], Fei Du[4,7], Chizuru Kawashima[1], Hideto Soeda[1], Satoshi Nakano[8], Yoshiya Uwatoko[4], Yutaka Ueda[4,9], Taku J. Sato[2], Kenya Ohgushi[4,10]

1. College of Humanities & Sciences, Nihon University, Sakurajosui, Setagaya-ku, Tokyo 156-8550, Japan
2. Institute of Multidisciplinary Research for Advanced Materials, Tohoku University, Sendai, Miyagi 980-8577, Japan
3. Institute for Materials Research, Tohoku University, Sendai, Miyagi 980-8577, Japan
4. Institute for Solid State Physics (ISSP), University of Tokyo, Kashiwa, Chiba 277-8581, Japan
5. Bragg Institute, Australian Nuclear Science and Technology Organisation, Locked Bag 2001, Kirrawee DC NSW 2232, Australia
6. School of Chemistry, The University of Sydney, Sydney, NSW 2006, Australia
7. Key laboratory of Physics and Technology for Advanced Batteries (Ministry of Education), College of Physics, Jilin University, Changchun 130012, People's Republic of China
8. Ultra-High Pressure Processes Group, National Institute for Materials Science, Tsukuba, Ibaraki 305-0044, Japan
9. Toyota Physical and Chemical Research Institute, Nagakute, Aichi 480-1192, Japan
10. Department of Physics, Tohoku University, Sendai, Miyagi 980-8578, Japan

1. **Neutron diffraction profiles of BaFe$_2$S$_3$ and their analysis**

Powder neutron diffraction measurements were performed using crushed crystals of BaFe$_2$S$_3$ sample 1. The obtained diffraction patterns are well fit based only on BaFe$_2$S$_3$, as shown in Fig. S1. Consistent with an earlier report[18], the crystal structure is precisely consistent with the orthorhombic space group *Cmcm*. Table S1 summarizes the structural parameters at $T = 295$ K determined by Rietveld refinement using the FullProf suite.[31]

Additional magnetic reflections appear in the data below 110 K and all magnetic peak positions can be indexed by a propagation wave vector, $\bm{q}_\mathrm{m} = (1/2, 1/2, 0)$. We employed group theory analysis to identify the magnetic structure that is allowed by symmetry. Basis vectors (BVs) of the irreducible representations (irreps) for the wave vector were obtained using the SARA*h* code.[32] There are four irreps in total, and each irrep consists of three BVs giving either parallel or antiparallel relationships between uniaxial magnetic moments along one crystallographic axis, as shown in Table S2. We initially sorted out all BVs by comparing $\chi^2$-factors and found $\psi_1$ had the best fit with $\chi^2 = 4.5$. The second best refinement was achieved using $\psi_2$ with 4.7, and the others poorly described our data ($\chi^2 > 8$). These two BVs have stripe-type magnetic structures, the only difference being the moment direction; $\psi_1$ along the *a* and $\psi_2$ along the *b*. We subsequently assumed that the magnetic structure could be represented by multiple BVs belonging to the same irrep and then tested the application of various combinations of BVs in $\Gamma_1$. The analysis with a combination of $\psi_1$ and $\psi_2$, however, did not improve the quality of the fit. Therefore we concluded that the magnetic structure of the material is of stripe type with magnetic moment direction along the *a* axis. This is consistent with the susceptibility data showing that the magnetic easy-axis is along the *a* axis. The data at $T = 4$ K together with the Rietveld refinement incorporating $\psi_1$ are depicted in Fig. S1.

2. **Specific heat of BaFe$_2$S$_3$**

Specific heat measurements were performed for BaFe$_2$S$_3$ sample 1 using a Physical Properties Measurement System (Quantum Design). As can be seen from Fig. S2, the specific heat (*C*) exhibits a small anomaly in the vicinity of $T_\mathrm{N}$. The entropy associated with the magnetic transition is estimated to be 0.051 J/K mol (see the right inset of Fig. S2), which is only 0.44 % of 2*R*ln2 (*R* being the gas constant). Such a small entropy release across the magnetic transition is consistent with the Mössbauer spectroscopy results showing a gradual increase in the volume fraction of the magnetically ordered phase below $T_\mathrm{N}$ (see the discussion below).

The $C$ curve in the low-temperature region was analysed by fitting the data below 4 K to the relationship $C = \gamma T + \beta T^3$, where the $\gamma$ and $\beta$ terms correspond to the electron and phonon contributions, respectively (left inset of Fig. S2). The resulting values were $\gamma = 0.38$ mJ/K$^2$ and $\beta = 0.79$ mJ/K$^4$. The negligible electron contribution is consistent with the insulating nature of the system. The $\beta$ coefficient is related to the Debye temperature, $\theta_D$, such that $\beta = (12\pi^4/5)NR/\theta_D^3$, where $N$ (= 6) is the number of atoms per unit cell. We calculate the $\theta_D$ value of 244 K, which is larger than the values previously reported for $A$Fe$_2$Se$_3$ ($A$ = Cs and Ba), owing to the lower atomic mass of S compared to Se[14,16]. In our analysis, we were unable to acquire useful information concerning magnons, which likely contribute to the low-temperature specific heat.

### 3. Mössbauer spectra of BaFe$_2$S$_3$.

[57]Fe Mössbauer spectra were acquired at temperatures down to 9 K, using crushed crystals of BaFe$_2$S$_3$ sample 1. The temperature evolution of the spectra is shown in Fig. S3a. At room temperature, the spectra can be well described by one quadrupole split paramagnetic lines. The isomer shift (IS) obtained by fitting the data with one doublet is IS = 0.44(1) mm/s, which is larger than the value of 0.39(1) mm/s obtained for CsFe$_2$Se$_3$ (ref. 16). If we apply Goodenough's empirical formula, IS = 1.68 − 0.5$m$ (ref. 30), we obtain an Fe valence of $m$ = 2.48 for BaFe$_2$S$_3$. This is much larger than the expected divalent state based on the assumption of Ba$^{2+}$ and S$^{2-}$ ionic states, and the reason for this discrepancy is unclear at present.

Below $T_N$, a magnetically split sextet gradually increases in intensity with decreasing temperature. A two-phase coexisting feature such as this, which is incompatible with the second-order phase transition, is also discernible in CsFe$_2$Se$_3$ (ref. 16). One plausible reason for this phenomenon is that $T_N$ is so sensitive to the microscopic parameters of the system in this low-dimensional structure that the spatial inhomogeneity in the magnetic state readily emerges. We note that this gradual evolution of the magnetic order is related to the small entropy release at $T_N$ observed in the specific heat. The magnetic hyperfine field (HF) deduced from the sextet gradually develops below $T_N$, and reaches a constant value of 13.5 T at 9 K, as shown in Fig. S3b. At the lowest temperature, the spectrum includes an additional contribution from the hyperfine field of 24.4 T, which is indicated by the blue coloration in Fig. S3a. This likely originates from traces of impurity phases such as Fe$_3$O$_4$.

### 4. Sample dependence of the magnetic and superconducting properties

Throughout this study, it was evident that the presence or absence of superconductivity under pressure greatly depends on the synthetic procedure. We therefore describe herein in detail the variations in the magnetic properties as well as the high-pressure resistivity of the various samples. The detailed sample synthesis procedure is provided in the Method section.

Figure S4 displays the magnetic susceptibility under a magnetic field of 5 T applied along the leg direction for samples 1 through 5. One can clearly see that the antiferromagnetic transition temperature ($T_N$) displays a significant dependence on the synthesis conditions, since $T_N$ values of 119, 105, 118, 122 and 124 K were obtained for samples 1 through 5, respectively. The sample 6 which is used for the resistivity measurement using the cubic anvil press shows the same $T_N$ as sample 5. Even though our chemical analysis could not detect any differences in the elemental ratios among these samples, we speculate that samples 1 and 2, fabricated from starting reagents without an excess of Fe, may be slightly Fe deficient. This Fe deficiency likely lowers the $T_N$ values, particularly in sample 2.

Even though the temperature dependence of the resistivity at ambient pressure does not exhibit an overly strong dependence on the synthetic procedure, a wide variation is seen in the resistivity values obtained under high pressure, as presented in Figs. S5a-5f. In all cases, the metal-insulator transition was observed around 10 GPa. In addition, superconductivity was observed for samples 1, 3, 4, 5 and 6, even though it was difficult to observe zero resistance for diamond anvil cell (DAC) measurements owing to the solid pressure-transmitting medium used in these measurements. The zero resistance is seen in sample 6 using cubic anvil press, as shown in Fig. 3e. The $T_c$ values also show a slight variation: the maximum $T_c$'s are 15, 15.5, 17.5 and 14 K for samples 1, 3, 4 and 5, respectively. Interestingly, we did not observe superconductivity in sample 2, likely because randomness due to the Fe deficiency disallows the superconducting state.

5. **Difference between high-pressure experiments using diamond anvil cell and cubic anvil press**

While we could clearly observe the zero resistance in high-pressure experiments using cubic anvil press, we could not observe zero resistance in high-pressure experiments using DAC. We here explain the reason. The difference is likely relevant to the non-hydrostatic pressure conditions generated in DAC using solid pressure transmitting medium, NaCl. The sample is compressed in one axial direction inside the pressure transmitting medium. We consider superconductivity of this material is very sensitive to external stress due to the one-dimensional nature, and non-hydrostatic condition easily deteriorates the superconducting paths. When the cubic anvil press is used, the sample is placed in a Teflon capsule with a liquid transmitting medium, glycerine. The Teflon cell was placed in a cubic-type gasket made of MgO. The cubic gasket is compressed equally in three axial directions. Nearly hydrostatic pressure can be realized even at low temperature and high pressure in it. Then, the zero resistance could be observed.

## 6. X-ray diffraction profiles under pressure

We performed X-ray diffraction analyses under high pressure at BL-18C at the Photon Factory and the resulting patterns are shown in Fig. S6. No definite structural phase transition was observed during these measurements, even at temperatures down to 10 K. This indicates that the superconductivity occurs while the material is in the ladder structure. The estimated lattice parameters are provided in Fig. 4a of the main text. The compressibility of the material was calculated as $\kappa = 19.7 \times 10^{-3}$ GPa$^{-1}$ and the linear compressibility values were determined to be $\kappa_a = 5.6 \times 10^{-3}$ GPa$^{-1}$ (rung direction), $\kappa_b = 11.5 \times 10^{-3}$ GPa$^{-1}$ (interlayer direction) and $\kappa_c = 3.1 \times 10^{-3}$ GPa$^{-1}$ (leg direction).

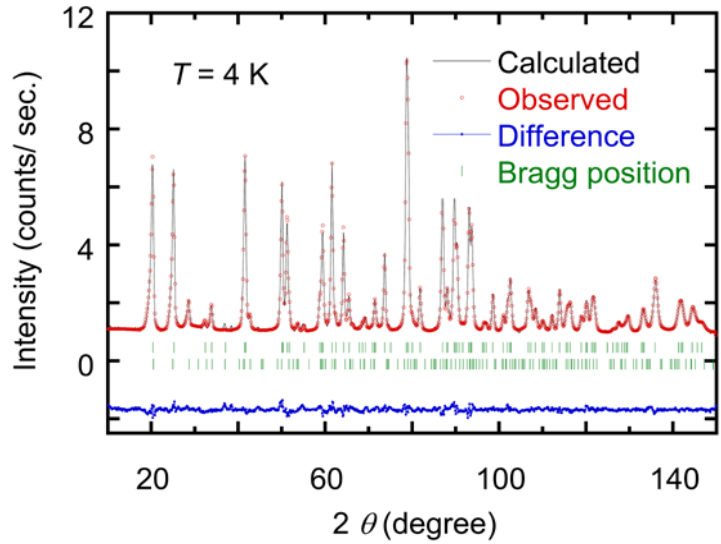

**Figure S1.** High-resolution neutron powder pattern at 4 K with Rietveld refinement results (solid lines). The calculated Bragg positions of nuclear and magnetic reflections are indicated by green ticks. The difference between the observed and calculated intensities is given by the blue line.

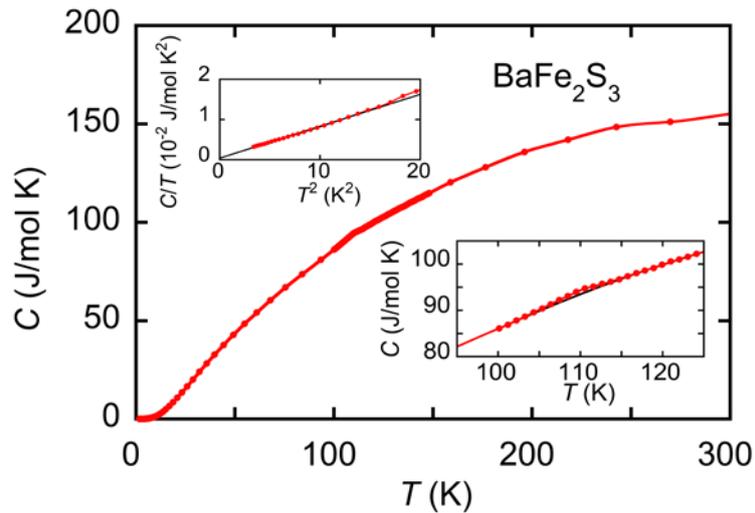

**Figure S2.** Temperature ($T$) dependence of the specific heat ($C$) of $BaFe_2S_3$. The right inset shows an expanded view around the antiferromagnetic transition temperature, in which the black line indicates the contribution of phonons. The left inset is an expanded view of the low-temperature region, in which the black line indicates the fitting results.

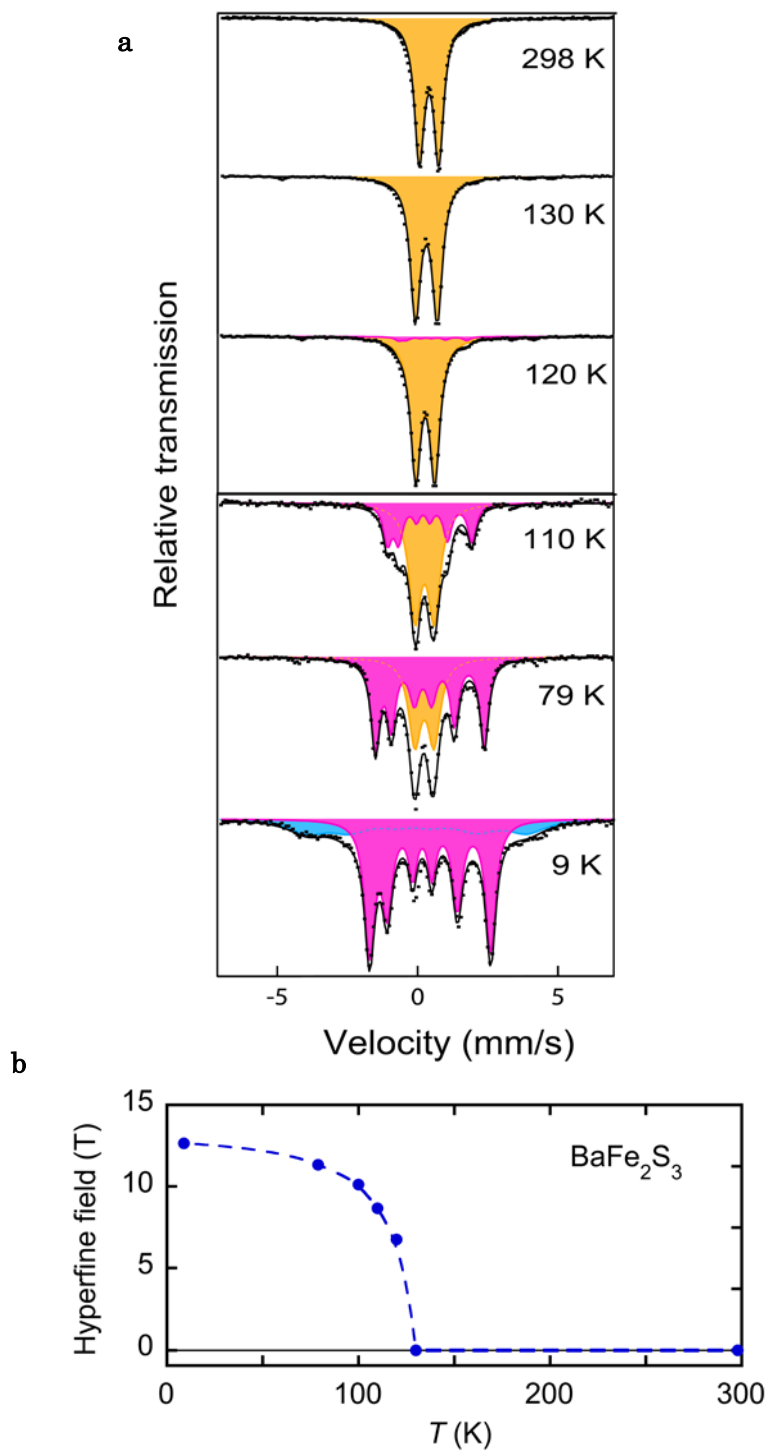

**Figure S3.** Mössbauer spectra of BaFe$_2$S$_3$. **a.** Temperature ($T$) dependence of the Mössbauer spectra, fitted with one doublet (yellow) and one sextet (pink). The 9 K spectra include an additional contribution (blue) originating from impurities. **b.** Temperature dependence of the hyperfine field.

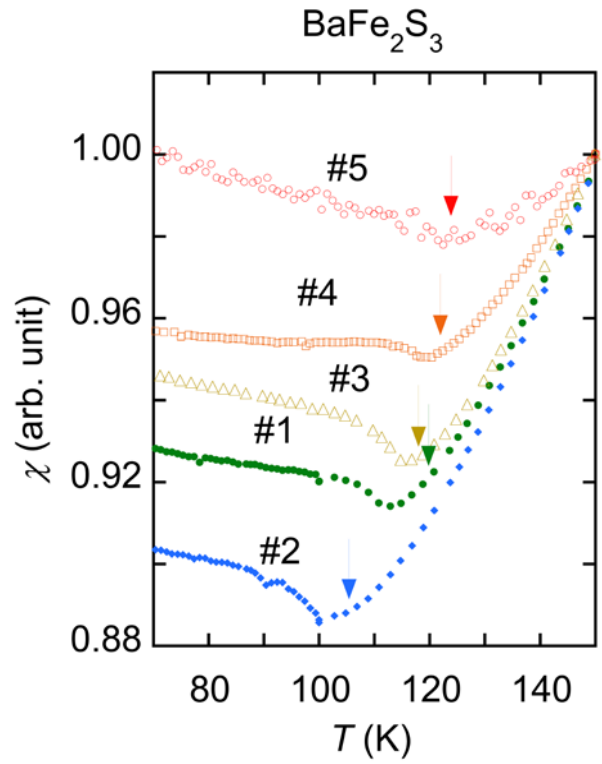

**Figure S4.** Temperature ($T$) dependence of the normalized magnetic susceptibility ($\chi$) for BaFe$_2$S$_3$ samples 1 through 5. The arrows indicate the antiferromagnetic transition temperature ($T_N$).

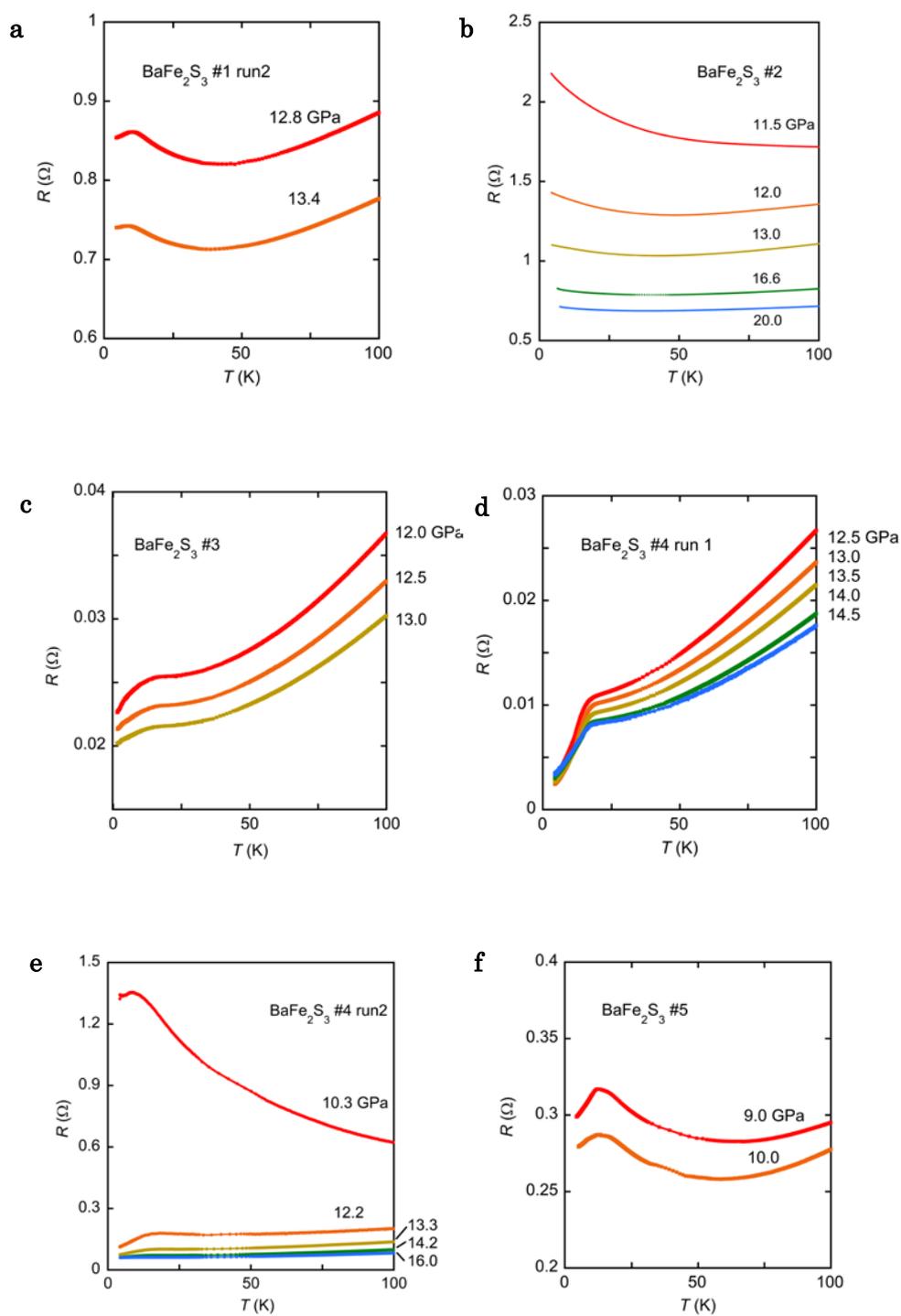

**Figure S5.** Temperature (*T*) dependence of the resistance (*R*) under high pressure for BaFe$_2$S$_3$, samples 1 through 5.

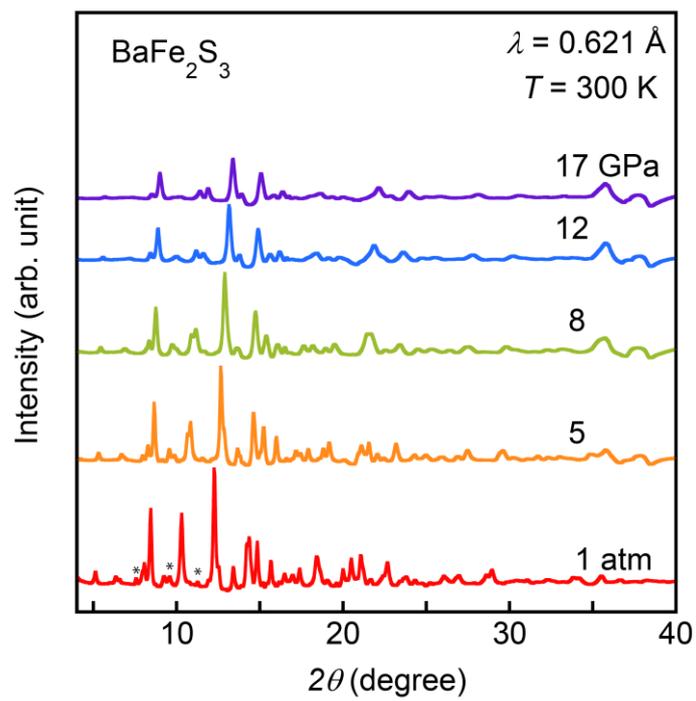

**Figure S6.** X-ray diffraction profiles under high pressure for BaFe$_2$S$_3$. Tiny impurity peaks are seen as designated by * symbols.

**Table S1.** Atomic positions within the space group *Cmcm* of $BaFe_2S_3$ at $T = 295$ K as determined by Rietveld analysis ($R_p = 3.45\%$).

| Atom | Site | $x$ | $y$ | $z$ | $B_{iso.}$ (Å$^2$) |
|------|------|-----|-----|-----|--------------------|
| Ba   | 4c   | 1/2 | 0.1845(5) | 1/4 | 1.54 (15) |
| Fe   | 8e   | 0.3461(3) | 1/2 | 0 | 1.43 (8) |
| S1   | 4c   | 1/2 | 0.6156(11) | 1/4 | 1.4 (2) |
| S2   | 8g   | 0.2074(7) | 0.3824(7) | 1/4 | 1.1 (2) |

**Lattice constants are $a = 8.787(1)$ Å, $b = 11.225(1)$ Å and $c = 5.288(1)$ Å. The isotropic Debye-Waller factor ($B_{iso.}$) was employed.**

**Table S2.** Basis vectors (BVs) of irreducible representations (irreps) for the space group *Cmcm* with the magnetic wave vector $q_m = (1/2,1/2,0)$.

| irrep | BV | #1 | #2 | #3 | #4 |
|-------|-----|----|----|----|-----|
| $\Gamma_1$ | $\psi_1$ ($\parallel a$) | 1 | -1 | 1 | -1 |
|  | $\psi_2$ ($\parallel b$) | 1 | -1 | 1 | -1 |
|  | $\psi_3$ ($\parallel c$) | 1 | 1 | 1 | 1 |
| $\Gamma_2$ | $\psi_4$ ($\parallel a$) | 1 | -1 | -1 | 1 |
|  | $\psi_5$ ($\parallel b$) | 1 | -1 | -1 | 1 |
|  | $\psi_6$ ($\parallel c$) | 1 | 1 | -1 | -1 |
| $\Gamma_3$ | $\psi_7$ ($\parallel a$) | 1 | 1 | 1 | 1 |
|  | $\psi_8$ ($\parallel b$) | 1 | 1 | 1 | 1 |
|  | $\psi_9$ ($\parallel c$) | 1 | -1 | 1 | -1 |
| $\Gamma_4$ | $\psi_{10}$ ($\parallel a$) | 1 | 1 | -1 | -1 |
|  | $\psi_{11}$ ($\parallel b$) | 1 | 1 | -1 | -1 |
|  | $\psi_{12}$ ($\parallel c$) | 1 | -1 | -1 | 1 |

**Columns legend: #1: ($x,1/2,0$), #2: ($-x +1,1/2,1/2$), #3: ($-x +1,1/2,0$) and #4: ($x,1/2,1/2$).**

simulated annealing and representational analysis (SARAh). *Physica* B **276,** 680-681 (2000).